\documentclass[twocolumn,twoside,slac_two]{revtex4}
\usepackage{graphicx}
\usepackage{fancyhdr}
\begin{document}

\title{MAGIC low energy observation of GRB\,090102 afterglow}

%

\author{A. Carosi, L.A. Antonelli}
\affiliation{ASI Science Data Center, Frascati, Italy and INAF National Institute for Astrophysics, I-00136 Rome, Italy}
\author{U. Barres de Almeida}
\affiliation{Max-Planck-Institut f\"ur Physik, D-80805 M\"unchen, Germany}
\author{D. Bastieri}
\affiliation{Universit\`a di Padova and INFN, I-35131 Padova, Italy}
\author{J. Becerra Gonz\'alez}
\affiliation{Depto. de Astrof\'{\i}sica, Universidad de La Laguna, E-38206 La Laguna, Spain}
\author{E. Colombo, M. Garczarczyk}
\affiliation{Inst. de Astrof\'{\i}sica de Canarias, E-38200 La Laguna, Tenerife, Spain}
\author{S. Covino, A. La Barbera, S. Spiro}
\affiliation{INAF National Institute for Astrophysics, I-00136 Rome, Italy}
\author{A. Dom\'{\i}nguez}
\affiliation{Inst. de Astrof\'{\i}sica de Andaluc\'{\i}a (CSIC), E-18080 Granada, Spain - Now at: Department of Physics and Astronomy, University of California,Riverside, CA 92521, USA}
\author{M. Gaug}
\affiliation{Universitat Aut\`onoma de Barcelona, E-08193 Bellaterra, Spain}
\author{F. Longo}
\affiliation{Universit\`a di Trieste, and INFN Trieste, I-34100 Trieste, Italy}
\author{V. Scapin}
\affiliation{Universidad Complutense, E-28040 Madrid, Spain}
\author{On behalf of the MAGIC Collaboration}

\begin{abstract}

Hints for a GeV component in the emission from GRBs are known since the EGRET observations during the ’90s and they have been recently confirmed by the data of the Fermi satellite. These results have, however, shown that  a fully satisfactory interpretative framework of the GRB phenomena is still lacking. The MAGIC telescope opens the possibility to extend the measurement of GRBs in the several tens up to hundreds of GeV energy range. From the theoretical point of view, both leptonic and hadronic processes have been suggested to explain the possible GeV/TeV counterpart of GRBs. Observations with ground-based telescopes of very high energy photons (E$>$30 GeV) from these sources are going to play a key role in discriminating among the different proposed emission mechanisms which are barely distinguishable at lower energies. MAGIC telescope observations of the GRB\,090102 (z=1.547) field from 03:14:52 UT to 06:54:01 UT are analyzed to derive upper limits to the GeV/TeV emission. We compare these results to the expected emissions evaluated for different processes in the framework of the standard fireball model. The results we obtained are compatible with the expected emission but cannot yet set further constraints on the theoretical scenario. However, the difficulty in modeling the low energy data for this event makes it difficult to fix in an unambiguous way the physical parameters which describe the fireball. Nonetheless, the MAGIC telescope, thanks to its low energy threshold and fast repositioning, is opening for the first time the possibility to fill the energy gap between space-based gamma detectors and the ground-based measurements. This will makes possible GRBs multiwavelength studies in the very high energy domain.

\end{abstract}

\maketitle

\thispagestyle{fancy}


\section{INTRODUCTION}

Since the discovery of Gamma-Ray Burst (GRB) afterglows in the late 90’s ~\cite{Costa97}, these energetic phenomena have been targets of observational efforts at essentially all electromagnetic wavelengths. The wealth of available information put severe constraints on the various families of interpretative scenarios, showing an unexpected richness and complexity of possible behaviors ~\cite{Geh09}. The first observations at MeV-GeV energies, with the Energetic Gamma-Ray Experiment Telescope (EGRET) on-board the Compton Gamma-Ray Observatory (CGRO) ~\cite{Hurley94}, showed that the observations in the high energy (HE: 1 MeV-30 GeV ) and in the very high energy range (VHE: 30 GeV - 30 TeV) can be powerful diagnostic tools for the emission processes and physical conditions of GRBs. The launch of \textit{Fermi} ~\cite{Band09}, with its Large Area Telescope (LAT) ~\cite{Atwood09}, showed that, at least for the brightest events, GeV emission from GRBs is a relatively common phenomenon ~\cite{Granot10} confirming that our comprehension of GRB physics is still unsatisfactory. In this context, multiple efforts have been performed in GRBs observations by the new generation of ground-based imaging atmospheric Cherenkov telescopes (IACTs), such as MAGIC, HESS or VERITAS. In all cases only upper limits have been derived.  The two most limiting factors are the  heavily attenuation of the VHE signal by pair-production due to the interactions with the lower energetic photons of the extragalactic background light (EBL) ~\cite{Nikishov62} and the delay of the observations. Within this scenario,  MAGIC has the advantage, compared to other IACTs, in its low energy sensitivity and pointing speed ~\cite{Garcz09}.  Throughout the paper the convention $Q_x = Q/10^x$ has been adopted in CGS units.

 \section{GRB\,090102}

GRB\,090102 was detected and located by the \textit{Swift} satellite ~\cite{Geh04} on January 02, 2009 at 02:55:45 UT ~\cite{Mangano09}. The prompt light curve is structured in four partially overlapping peaks for a total $T_{90}$ of $27.0 \pm 2.0 $ s. Since the burst was also detected by \textit{Konus Wind} ~\citep{Golenetskii09} and \textit{Integral} ~\cite{Mangano09b}, it has been possible to obtain a very good reconstruction of the prompt emission spectral parameters. The time-averaged spectrum can be modeled with the classical Band function ~\cite{Band93} with peak energy $E_{peak}=451^{ \ +73}_{ \ -58}$ keV and a total fluence in the 20 keV - 2 MeV range of $3.09^{ \ +0.29}_{ \ -0.25} \times 10^5$ erg cm$^{-2}$ ~\cite{Golenetskii09}. Early Optical follow-up measurements were performed by many groups like TAROT ~\cite{Klotz09} at $T_0$+40.8 s, the REM robotic telescope at $T_0$+53 s ~\cite{Covino09} and GROND telescope ~\cite{Afonso09} at $T_0$+2.5 h. The optical light-curve, monitored from several tens of seconds to slightly more than a day from the $T_0$, shows a steep-to-shallow behavior with a break at about 1ks. Before the break, the optical flux decay index is $\alpha_1=1.50 \pm 0.06$ while the index became $\alpha_2=0.97 \pm 0.03$ after the break, steeper and flatter respectively, when compared to the simultaneous X-ray emission. Optical spectroscopy was rapidly carried out by the NOT telescope ~\cite{deug09} deriving a redshift of z=1.547, which allows to compute an isotropic energy value of $E_{iso}=5.75 \times 10^{53}$ erg and a rest frame peak energy of $E_{peak}$=1149$^{ \ +186}_{ \ -148}$ keV in good agreement with the Amati relation. A detailed discussion of the follow up observations for this burst can be found in ~\cite{Gendre10}.

\section{MAGIC OBSERVATION}

The MAGIC telescope located at Roque de los Muchachos (28.75$^{\circ}$N, 17.89$^{\circ}$W, La Palma, Canary Islands) performed a follow-up measurement of GRB\,090102 starting the observation at the position provided by {\it Swift} (RA: 08h 33m 02s; DEC: 33$^{\circ}$ 05' 29'' ) at 03:14:52 UT at $T_0+1161$ s. Data presented in this paper have been taken before November 2009, when MAGIC was operating as a single telescope. First data runs were taken at very low zenith angles from 5$^{\circ}$ reaching 52$^{\circ}$ at the end of datataking at 06:54:01 UT after 13149 s of observation. MAGIC upper limits above 80 GeV have already been published for this GRB ~\cite{Gaug09} while results and scientific discussion about a successive dedicated analysis focused in the low energy band ~\cite{Gaug09b} will be presented here. To ensure the lowest energy threshold, only data taken with zenith distance $<$ 25$^{\circ}$, corresponding to the first 5919 s of observation (data sub-sample up to 04:53:32 UT) has been taken into account during this analysis. Together with the MAGIC-1 sum trigger system ~\cite{Albert08}, it yields an analysis threshold around 30 GeV evaluated using MC simulations. In order to accurately estimate the background from hadronic atmospheric showers, an OFF data sample has been taken one night later with the telescope pointing close to the burst location and in the same observational conditions and instrument setup. In spite of the excellent MAGIC low energy analysis threshold, no significant excess of gamma-ray photons has been detected from a position consistent with GRB\,090102. Differential upper limits with 95$\%$ CL were evaluated. Telescope efficiency was evaluated with a 30$\%$ estimation of systematic uncertainties. However, the excellent low energy threshold for ground-based observations with Cherenkov telescopes makes these observations particularly interesting in spite of the high redshift of GRB\,090102.

\begin{table}
\begin{center}
\begin{tabular}{|c|c|c|}\hline
{\bfseries E bin} &  {\bfseries Fluence Limits} & {\bfseries Average Flux Limits} \\
$[{\rm GeV}]$ & $[\rm{erg \ cm}^{-2}]$ & $[{\rm erg \ cm}^{-2} \ {\rm s}^{-1}]$\\
\hline
25 - 50 & $5.2 \times 10^{-6}$ & $8.7 \times 10^{-10}$ \\
50 - 80 & $8.9 \times 10^{-7}$ & $1.5 \times 10^{-10}$ \\
80 - 125 & $1.8 \times 10^{-6}$ & $3.1 \times 10^{-10}$ \\
125 - 175 & $1.3 \times 10^{-6}$ & $2.2 \times 10^{-10}$ \\
175 - 300 & $9.5 \times 10^{-7}$ & $1.6 \times 10^{-10}$ \\
300 - 1000 & $1.8 \times 10^{-7}$ & $0.3 \times 10^{-10}$ \\
\hline
\end{tabular}
\end{center}
\caption[UL Table]{MAGIC-I 95$\%$ confidence level upper limits for the afterglow emission of GRB\,090102. The values correspond to the first 5919 s of observation from 03:14:52 UT to 04:53:32 UT for a zenith distance ranging between 5$^{\circ}$ and 25$^{\circ}$. }
\label{tab:ul} 
\end{table}

\section{MODELING THE VHE EMISSION}

In the so-called standard scenario, known as the fireball model ~\cite{Pacz86}, GRB dynamics during the prompt phase are governed by relativistic collisions between shells of plasma emitted by a central engine (internal shocks). Similarly, the emission during the afterglow is thought to be connected to the shocks between these ejecta with the external medium (external shocks). In both cases, the observed photons are radiation from particles accelerated to ultrarelativistic energies by successive collisions with magnetized medium (Fermi mechanism). Even if the details of the acceleration process are not firmly known, several non-thermal mechanisms have been suggested to be sources of VHE photons. Possible processes comprise both leptonic (e.g. electron synchrotron, Synchrotron Self Compton-SSC) and hadronic models (e.g. proton synchrotron)~\cite{Zhang01}. In the most plausible scenario, electron synchrotron radiation is the dominant process in the low energy regime ($<$ MeV) while SSC emission overwhelm other processes in the VHE range. As first approximation, a broken power-law can be used in modeling the expected emission. The relevant break energies of the spectrum will be the minimum injection $\nu_m$ and the cooling $\nu_c$ . The first one refers to emission frequency of the bulk of the electrons population while the cooling frequency identifies where electrons effectively cool. Both are strongly dependent on the microphysical parameters used to describe the fireball and for the SSC are (following ~\cite{Zhang01}):

\begin{eqnarray}
\nu_{m}^{ssc} &=& 1.3 \times 10^{22} \ \ \epsilon_e^4  \left( \frac{p-2}{p-1} \right)^4 \nonumber \\
                                &\times& E_{52}^{3/4} n^{-1/4} t_h^{-9/4} \epsilon_B^{1/2} (1+z)^{5/4} \ [{\rm Hz}]
\end{eqnarray}
\begin{eqnarray}
\nu_{c}^{ssc} &=& 1.2 \times 10^{25} (1+Y_e)^{-4} \nonumber \\
                                &\times& E_{52}^{-5/4} n^{-9/4} \epsilon_B^{-7/2}  t_h^{-1/4} (1+z)^{-3/4} \ [{\rm Hz}]
\end{eqnarray}

where $n$ the medium particle density, $E$ is the energy per unit solid angle, $t_h$ the time delay after the GRB onset (in hours) and $z$ is the source redshift.  The slope of the electron energy distribution ($p$) can be evaluated using the optical spectral index light-curve ~\cite{Gendre10} and has been found to be $p=2.29 \pm 0.04$ in good agreement with numerical simulations which suggest a value of $p$ ranging between 2.2-2.3 ~\cite{Vie03}. On the contrary, we can only constrain the values for the micro-physical parameters $\epsilon_e$, the fraction of total energy going to electrons, and $\epsilon_B$, the fraction of total energy going to magnetic fields. Assuming that the optical light curve time break ~\cite{Melandri10} is less then the start time of the shallow decay phase ($T_{break} \lesssim  10^3$ s), we obtain $ 0.04 \lesssim \epsilon_{e} \lesssim 0.2$ and $ 7 \times 10^{-4} \lesssim \epsilon_B \lesssim 0.05$ which only barely fix the $\epsilon_B$, $\epsilon_e$ values. We thus assume $\epsilon_B \sim 0.01$ and $\epsilon_e \sim 0.1$ which correspond to typical values for the late afterglow ~\cite{Yos03}.\\

\begin{figure}
 \centering
 \includegraphics[width=\columnwidth,angle=270,width=82mm]{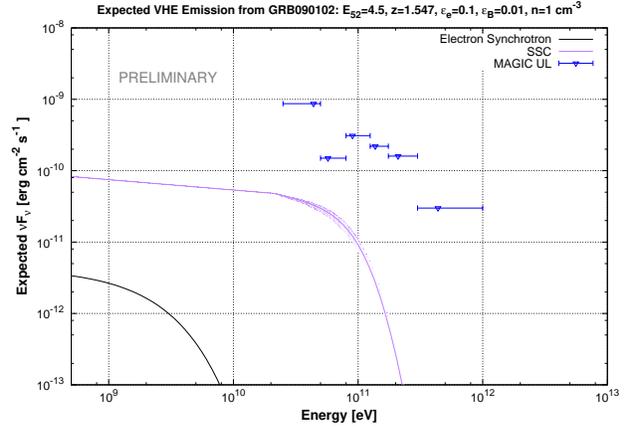}
 \caption{Expected SSC emission during the afterglow of GRB\,090102. Blue triangles are 95\% CL upper limits derived by MAGIC. The effect of the EBL absorption has been taken taking into account using the model in ~\cite{Alberto11}. The shaded region shows the uncertainty in the EBL absorption.}
 \label{fig:ul}
\end{figure}

\begin{figure}
 \centering
 \includegraphics[width=\columnwidth,angle=270,width=82mm]{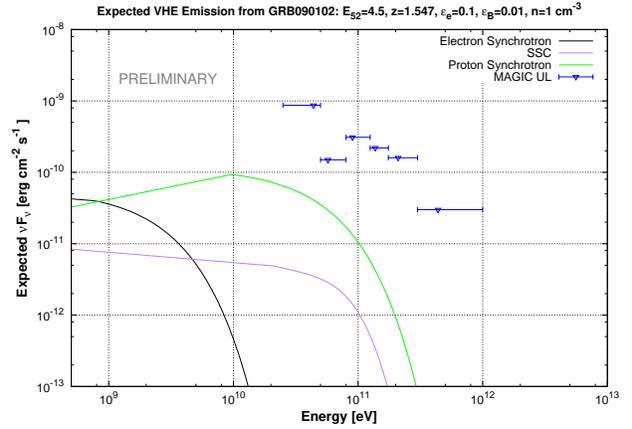}
 \caption{The hadronic component exceed the leptonic mechanism only for particular and non realistic choice of the microphysical parameters. However the cooling frequency for protons can easily reach the very high energy band. }
 \label{fig:ul2}
\end{figure}

\subsection{EBL absorption}

Gamma ray absorption by pair production with Extragalactic Background Light (EBL) limits the IACTs capability in detecting sources at redshift $z>$1. We adopted the model in ~\cite{Alberto11} to evaluate the EBL absorption. The EBL intensity evaluated using this model matches the minimum level allowed by galaxy counts, which leads to the highest transparency of the universe to VHE $\gamma$-rays. This is indeed an optimistic yet realistic scenario for GRBs studies in the VHE regime. As a matter of fact, we have obtained  a value for the optical depth $\tau$ of $0.218^{\ +0.075}_{\ -0.041}$ at about 40 GeV. This gives an attenuation of the flux at the same energy of the order of $\sim20\%$, a value that does not significantly compromise the detection capability of MAGIC. 

\section{DISCUSSION}

We have used the above derived equations in order to predict the expected emission in the MAGIC energy range. From numerical results it is evident that, for both the chosen parameters and at the MAGIC observation time, leptonic components are usually the dominant mechanisms from the low (electron synchrotron emission) to the very high energy (SSC). MAGIC observations ($>$ 40 GeV) were carried out in the SED region where $\nu$F$_{\nu} \propto \nu^{(2-{\rm p})/2}$ so that it is possible to evaluate the expected SSC emission:
\begin{eqnarray}
\nu F_{\nu} \propto \nu_{c,ssc}^{1/2}  \nu_{m,ssc}^{(p-1)/2} \nu^{(2-p)/2}
\end{eqnarray}
which gives (for our first energy bin and taking into account the EBL absorption) $\nu F_{\nu, ssc}$(40 GeV)$ \approx 3 \times 10^{-11}$ erg cm$^{-2}$ s$^{-1}$. 
This result is below of about one order of magnitude (in the first energy bins) to the reported upper limits. However, a change in the microphysical fireball values can influence the VHE emission giving scenarios with substantially higher flux. We also explored the possibility of a strong and prevalent proton synchrotron component in the GeV range. The cooling frequency for protons can in fact easily reach the GeV regime since $\frac{\nu_{c,p}}{\nu_{c,e}} \propto \left( \frac{m_{p}}{m_{e}} \right)^6$ . However, the required energy budget ($\sim 10^{55}$ erg) and medium density ($\sim 100$ cm$^{ -3}$) needed to match the expected low energy flux with observed data is rather incompatible with the measured total energy release. This makes the possibility to observe the hadronic emission component with the MAGIC telescope unrealistic, at least for a canonical fireball. Other hadronic-induced emissions such as $\pi^0$ decay \citep{BoDe98} can have a non negligible effect with new features in the spectrum in a higher energies range. Moreover, at these energies, $\nu F_{\nu} \propto t^{-1.1}$, implying that lowering the temporal delay of the observations can make the expected emission higher by one order of magnitude. Our estimates show that MAGIC follow-up observations within few hundreds of seconds from the T$_0$ would have the potential to detect this source. This demonstrates both the capabilities of the system and the necessity of a fast-response observations. Although we could not yet set firm constraints on the main theoretical scenarios, our results show the interesting perspectives for an afterglow detection in very high energy domain. This energy range will have very high impact on the understanding of the GRB phenomena and the unique opportunity of having simultaneous follow-up with \textit{Fermi}/LAT and the MAGIC telescope will have an important role in constraining different emission mechanisms and the space parameters.\\
The above results will be discussed in a forthcoming dedicated publication together with the FERMI/LAT data.

\begin{figure}
 \centering
 \includegraphics[width=\columnwidth,angle=270,width=82mm]{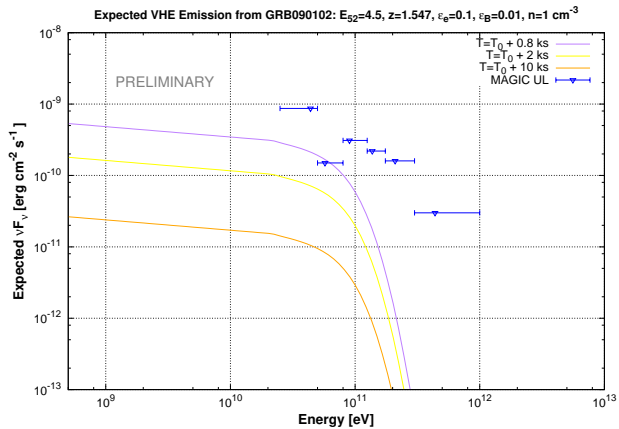}
 \caption{Expected SSC emission from GRB\,090102 at different time: T$_0$+0.8 ks (purple curve), T$_0$+2 ks (yellow curve), T$_0$+10 ks (orange curve).}
 \label{fig:ul2}
\end{figure}

\bigskip 
\begin{acknowledgments}
We would like to thank the Instituto de Astrofisica de Canarias for the excellent working conditions at the Observatorio del Roque de los Muchachos in La Palma. The support of the German BMBF and MPG, the Italian INFN and Spanish MICINN is gratefully acknowledged. This work was also supported by ETH Research Grant TH 34/043, by the Polish MNiSzW Grant N N203 390834, and by the YIP of the Helmholtz Gemeinschaft.\\
\end{acknowledgments}

\bigskip 

\end{document}